\documentclass[11pt]{article}

\usepackage{geometry}
\usepackage{asp2006}
\usepackage{graphicx}
\usepackage{amsmath}
\usepackage{subfigure}
\usepackage[symbol]{footmisc}
\def\ms{\rm\, m\;s^{-1}}
\def\kms{\rm\,km\;s^{-1}}

\def\H{\mathcal H}

\def\photon{{\textrm{\thinspace null}}}

\begin{document}
\title{Testing GR with Galactic-centre Stars}
\author{Raymond Ang\'elil\altaffilmark{1} and Prasenjit Saha\altaffilmark{1}}
\altaffiltext{1}{Institute for Theoretical Physics, University of Z\"urich, \\
Winterthurerstrasse 190, CH-8057 Z\"urich, Switzerland}  

\begin{abstract}
  The Galactic Centre S-stars orbiting the central supermassive black
  hole reach velocities of a few percent of the speed of light. The
  GR-induced perturbations to the redshift enter the dynamics via two
  distinct channels. The post-Newtonian regime perturbs the orbit from
  the Keplerian (Zucker et al., 2006, Kannan \& Saha 2009), and the
  photons from the Minkowski (Ang\'elil \& Saha 2010). The inclusion
  of gravitational time dilation at $\mathcal{O}\left(v^2\right)$
  marks the first departure of the redshift from the line-of-sight
  velocities. The leading-order Schwarzschild terms curve space, and
  enter at $\mathcal{O}\left(v^3\right)$. The classical Keplerian
  phenomenology dominates the total redshift. Spectral measurements of
  sufficient resolution will allow for the detection of these
  post-Newtonian effects. We estimate the spectral resolution required
  to detect each of these effects by fitting the redshift curve via
  the five keplerian elements plus black hole mass to mock data. We
  play with an exaggerated S2 orbit - one with a semi-major axis a
  fraction of that of the real S2. This amplifies the relativistic
  effects, and allows clear visual distinctions between the
  relativistic terms. We argue that spectral data of S2 with a
  dispersion $\sim 10\kms$ would allow for a clear detection of
  gravitational redshift, and $\sim 1\kms$ would suffice for
  leading-order space curvature detection.
 \end{abstract}

\section{Introduction}
The stars orbiting the supermassive black hole ($M\approx 4.4 \cdot
10^6 M_\odot$) within the central arcsecond are on highly relativistic
orbits. In comparison, the velocity of a geosynchronous Earth
satellite is $v_{\rm satell} \approx 0.00005c$. Mercury, whose orbital
Schwarzschild precession has been measured has $v_{\rm merc} \approx
0.00016c$ . Binary pulsar systems manage to reach $v_{\rm binary\
  pulsar} \sim 0.003c$, while the galactic centre S-Stars boast
$v_{\rm S2} \sim 0.03c$. This, due to their proximity to the black
hole during pericenter passage (down to $\sim 3000$ of the
gravitational radius) make this class of stars the fastest resolvable
ballistic objects known, and allow for the prospect of detecting
post-Newtonian effects.

The dynamics of the orbit and light trajectories provide an
opportunity to test the form of the metric, and in doing so, General
Relativity.  The Kerr metric is the external solution to the Einstein
Field Equations for a rotating body. The geodesics of such a
space-time exhibit some well-known features: gravitational time
dilation, prograde precession, lensing, and
frame-dragging. Gravitational time dilation and precession are orbital
effects. The former due to a temporal stretching, and the latter due
to space curvature. Such a curvature also affects photon trajectories
--- veering the trajectories away from those of straight lines. Each
of the these effects contributes to the redshift of the
Galactic-centre stars in a distinct manner.

These features, although markedly distinct, are all due to the same
metric. The post-Newtonian formalism, valid provided $r\gg0$, allows
us to cleanly disentangle these effects, and investigate the detection
of each separately. Not only are the stars on post-Keplerian orbits,
but the photons must travel through spacetime on nontrivial paths
before arriving at Earth. This nontrivial path through spacetime
affects the time of arrival, and therefore the redshift.

\vskip-24pt\null

\section{Model}
The star's orbit can be described by the Hamiltonian (Ang\'elil \&
Saha 2010)
\begin{equation}\label{timelikemetric}
\begin{array}{c l l l}
 \displaystyle \H_{star} = &\displaystyle -\frac{p_t^2}{2} &\displaystyle \propto v^1 & \mbox{No gravity} \\
 &\displaystyle  +\frac{p_r^2}{2} + \frac{p_\theta^2}{2r^2}+ 
\frac{p_\phi^2}{2r^2\sin^2\theta}-\frac{p_t^2}{r}  &\displaystyle \propto v^1, v^2 & \mbox{Kepler + Time-dilation}\\
 &\displaystyle -\frac{p_t^2}{r^2} - \frac{p_r^2}{r}&\displaystyle \propto v^3 & \mbox{space curvature} \\
 &\displaystyle + \mbox{ frame dragging, torquing, ...} & &
\end{array}
\end{equation}

At leading order, the system is spatially invariant, and the star
feels no acceleration. At next-to-leading order, gravity
debuts. Classically, the potential term is $1/r$. GR however demands
the modification to $p_t^2/r$, which results in gravitational
time-dilation $\propto v^2$, a consequence of the Einstein Equivalence
Principle. Spatially the problem has remained unchanged. However,
because the time-dilation term affects the photon arrival times, the
redshift is affected. Space curvature enters one order higher. This is
the leading-order Schwarzschild term, and causes the orbit to
precess. Higher order effects, such as frame-dragging and torquing, we
choose not to delve into here.

The Hamiltonian governing photon paths, being null, contains a
different selection of pre-truncation terms.

\begin{equation}\label{nullmetric}
\begin{array}{c l l l}
\displaystyle\H^\photon =  &  \displaystyle-\frac{p_t^2}{2}+\frac{p_r^2}{2}+\frac{p_\theta^2}{2r^2}+
\frac{p_\phi^2}{2r^2\sin^2\theta} &\displaystyle \propto v^0 & \mbox{Minkowski} \\
& \displaystyle-\frac{p_t^2}{r}-\frac{p_r^2}{r}&\displaystyle \propto v^3 & \mbox{space curvature}\\
 &\displaystyle + \mbox{ frame dragging, torquing, ...} & &
\end{array}
\end{equation}
At leading order, the trajectories are straight lines. Lensing occurs
at $\mathcal{O}\left(v^3\right)$ via the leading order Schwarzschild
contribution. There is no contribution at
$\mathcal{O}\left(v^4\right)$. For photons, the frame-dragging term
debuts at $\mathcal{O}\left(v^5\right)$ along with higher-order
Schwarzschild terms, and spin-induced torquing terms. In this work, we
consider effects only up to $\mathcal{O}\left(v^3\right)$ for both the
null and timelike cases\footnote{For a maximally spinning black hole,
  the frame-dragging \textit{photon} signal on S2's redshift at
  pericenter is $\sim 10 \ms$ --- two orders of magnitude weaker than
  the Schwarzschild photon signal. It is unlikely
  that the next generation of instruments will possess the capability
  to probe such deeper, weaker terms. We feel morally obligated not to
  raise the heartbeat of the observer reading this.}.

\vskip-24pt\null

\section{Calculating the redshift}
\begin{figure}[!ht]
\centering
\subfigure{
\includegraphics[height=2.0in, width = 1.3in,clip = true, trim = 50pt 50pt 115pt 95pt]{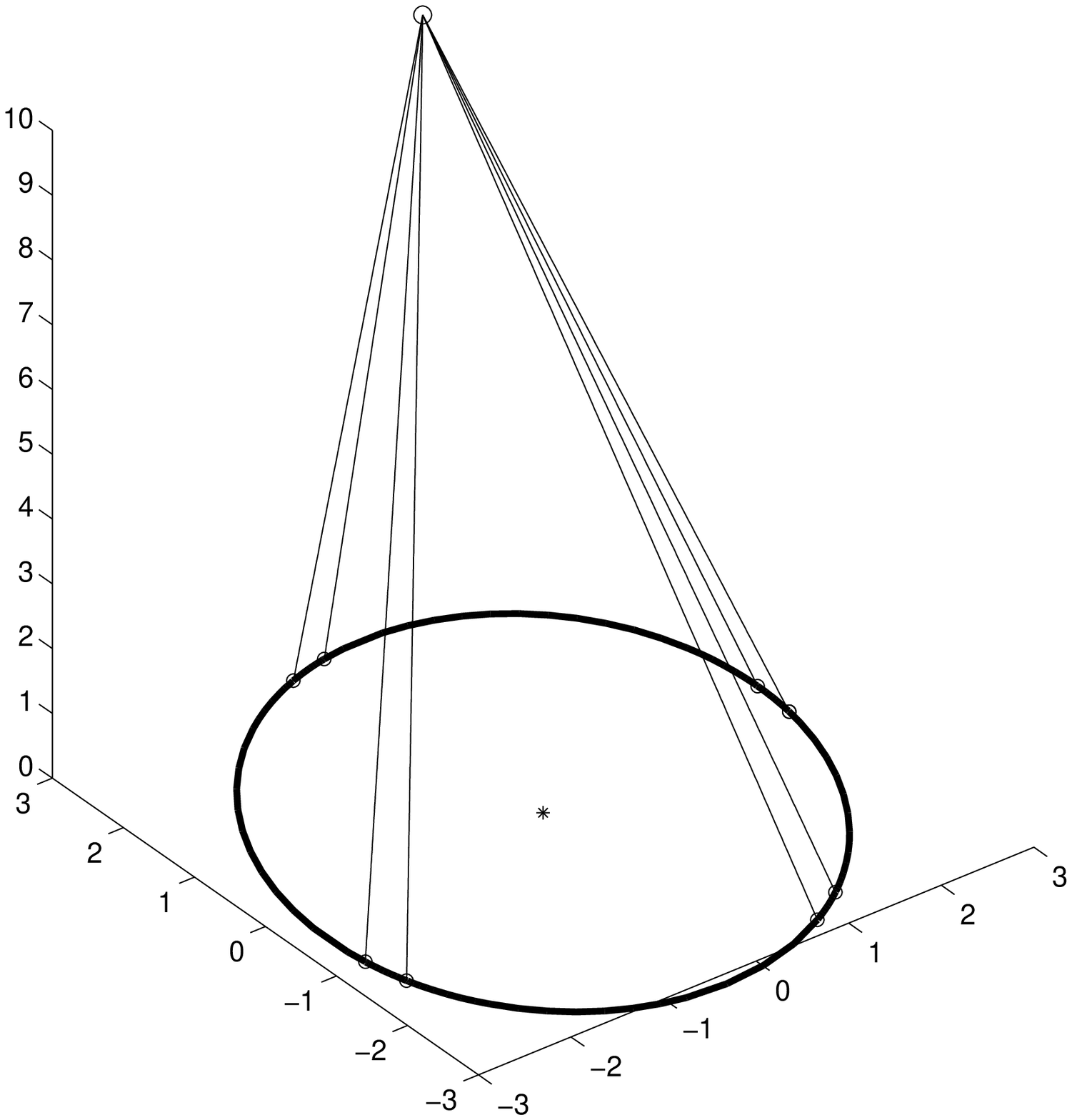}}
\subfigure{
\includegraphics[height=2.0in, width = 1.3in,clip = true, trim = 50pt 50pt 115pt 95pt]{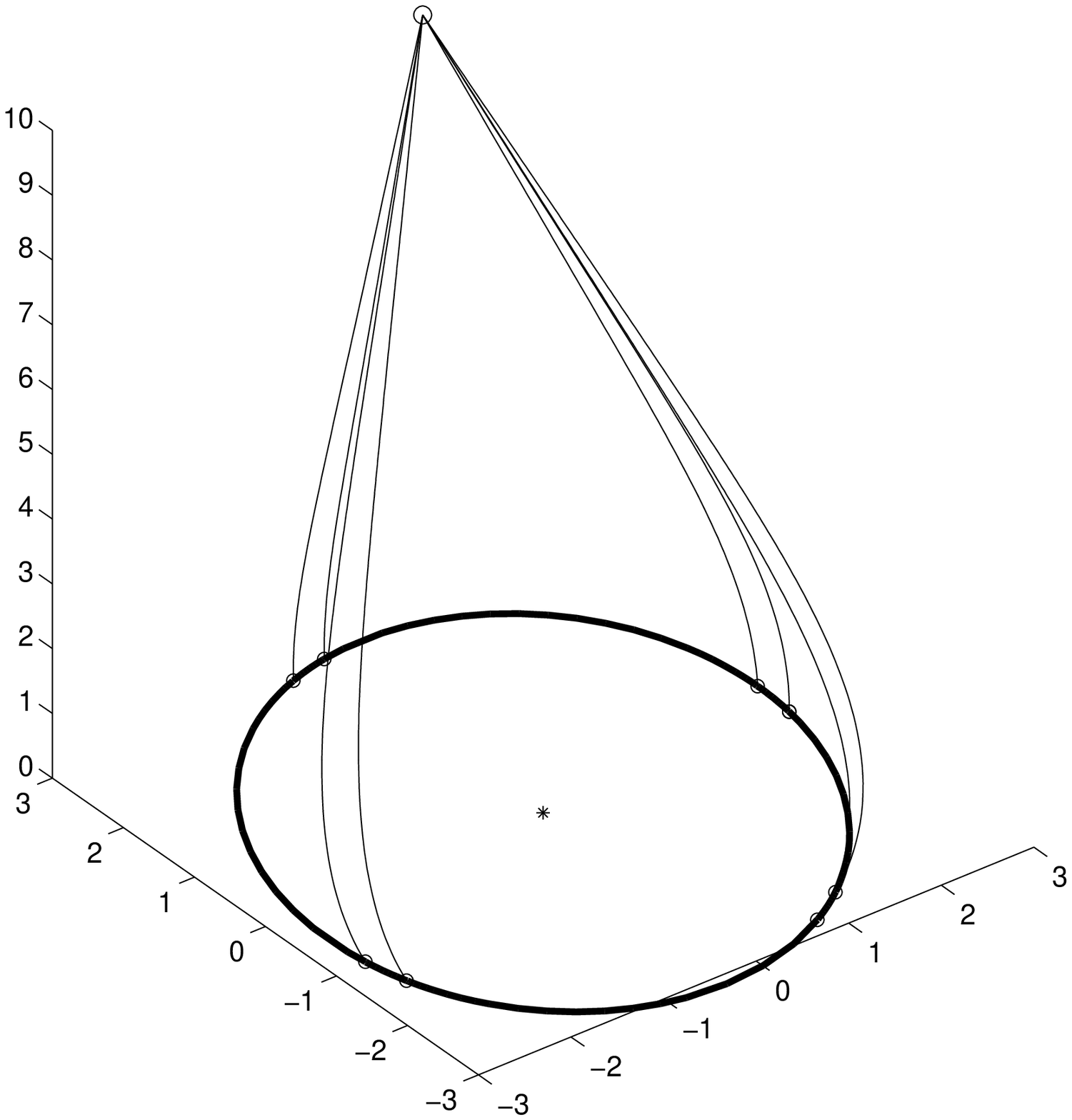}}
\subfigure{
\includegraphics[height=2.0in, width = 1.3in, clip = true, trim = 50pt 50pt 115pt 95pt]{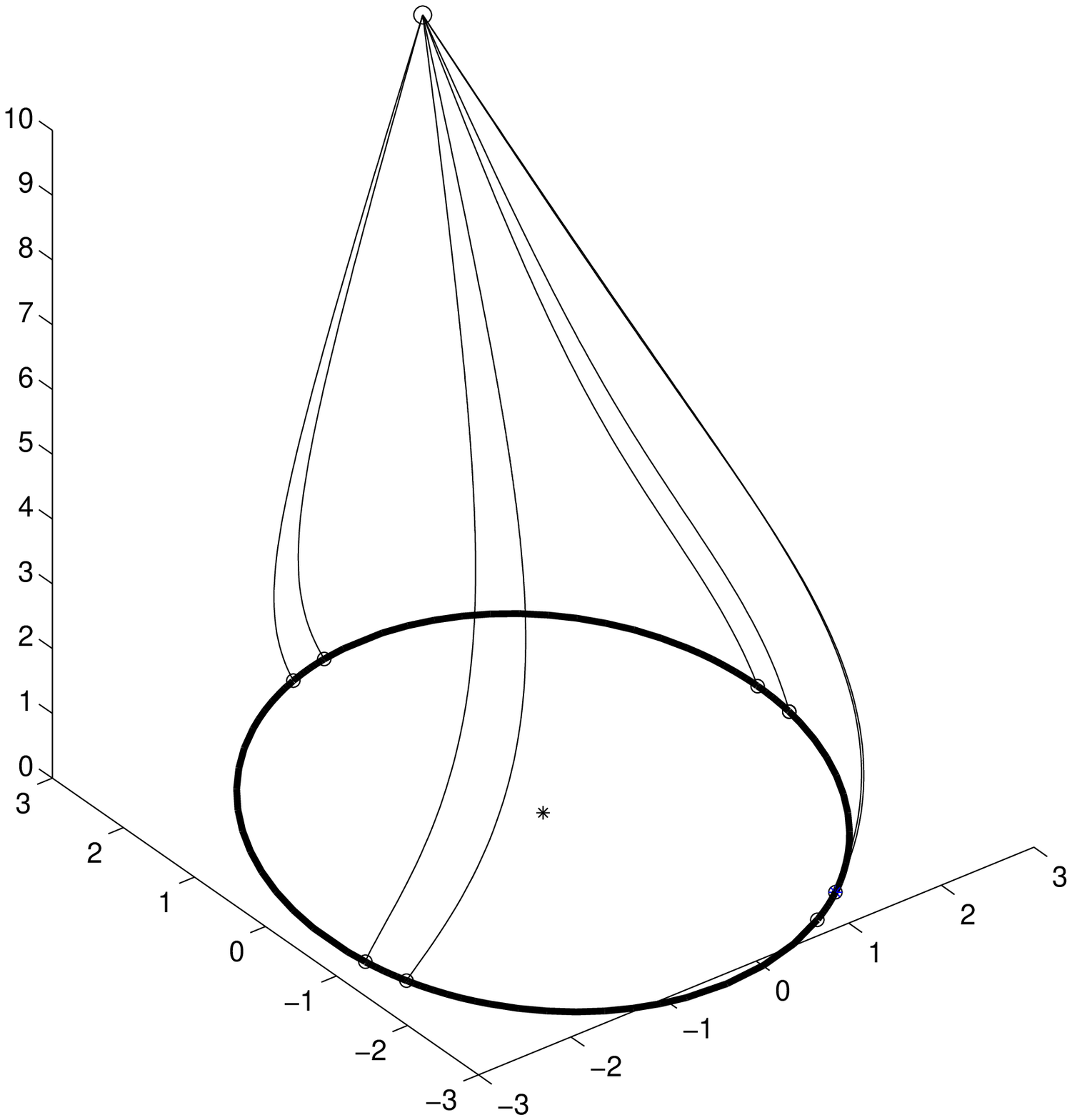}}
\caption{Schematic illustration of the redshift calculation method. Each photon shot by the star hits the observer. Each of the above cases yields four points on the redshift curve. The first panel shows Minkowski photons, the second space-curved photons, and the third frame-dragged photons.}\label{fig:method}
\end{figure}

To calculate the redshift curve of the star, we integrate the star's
orbit using timelike solutions to (\ref{timelikemetric}), and then, on
chosen points along the star's orbit, we find the paths of
\textit{those particular photons emitted by the star which hit the
 observer} (Figure \ref{fig:method}).  To do this, the initial angular
momentum of the trajectories (corresponding to null solutions of
\ref{nullmetric}) is varied until the termination position of the
photon converges on the observer position. Once the trajectories of
these photons are known, the redshift may then be calculated directly
from the definition:

\begin{equation}
z = \frac{t_{a_2} - t_{a_1}}{\tau_{e_2}-\tau_{e_1}}-1,
\end{equation} 
where $\tau_{e_1}$ and $\tau_{e_2}$ are the proper times of a pair of photons emitted at neighbouring points on the star's orbit, and $t_{a_1}$ and $t_{a_2}$ are their respective arrival times. 

\vskip-24pt\null

\section{Post-Newtonian detection}
\begin{figure}[!ht]
\centering
\subfigure{
\includegraphics[height= 4.2cm]{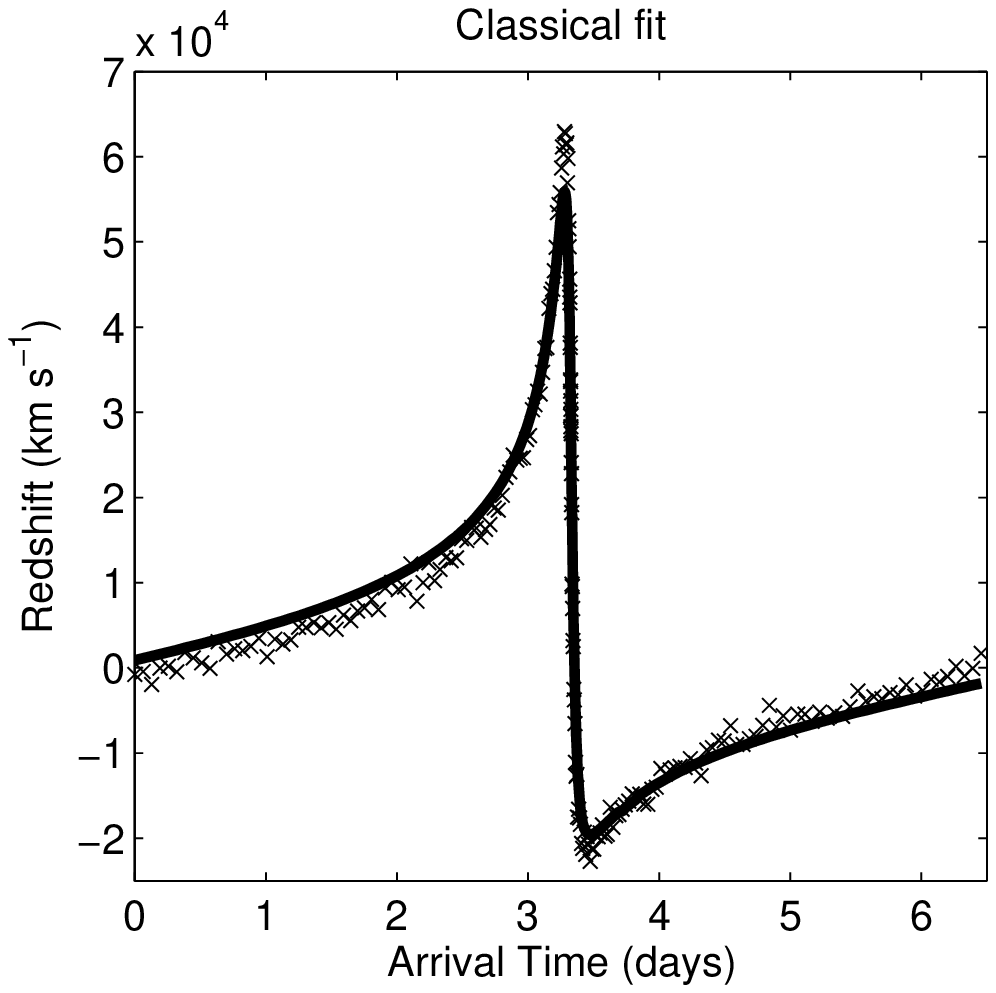}}
\subfigure{\includegraphics[height= 4.2cm]{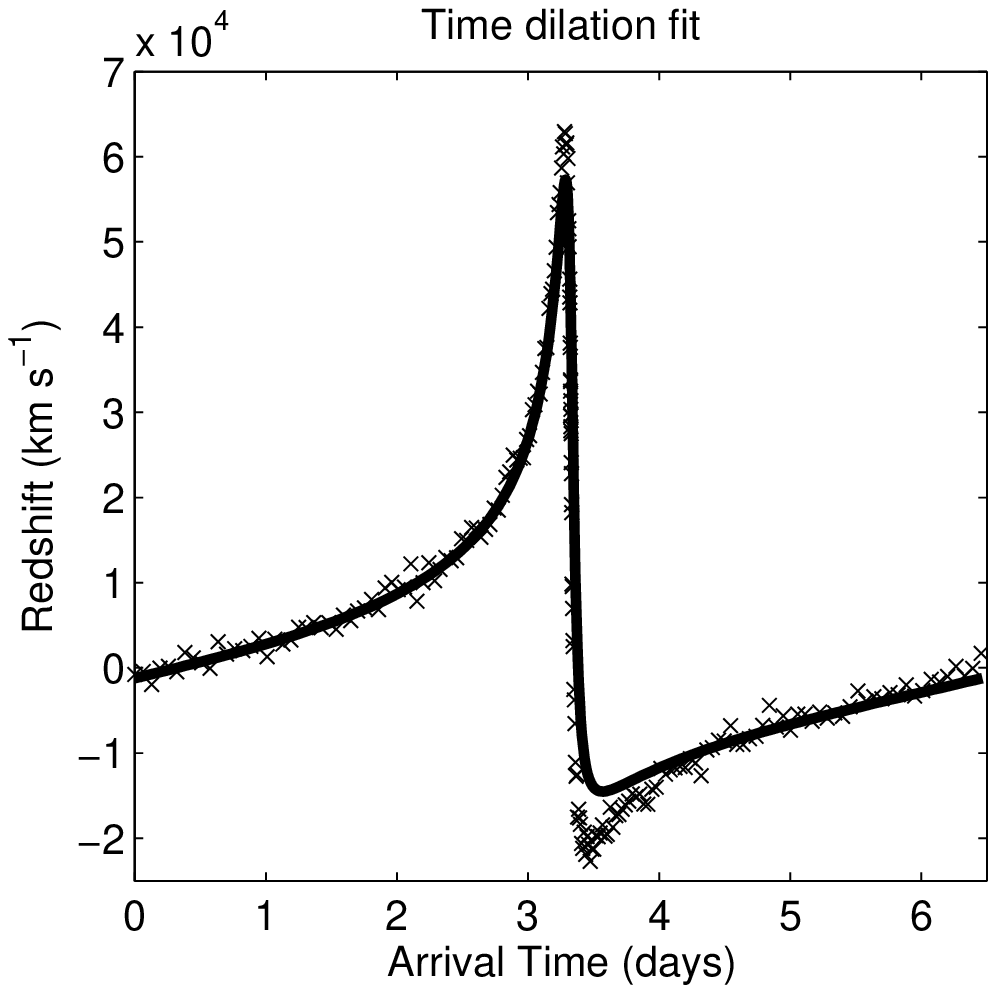}}
\subfigure{\includegraphics[height= 4.2cm]{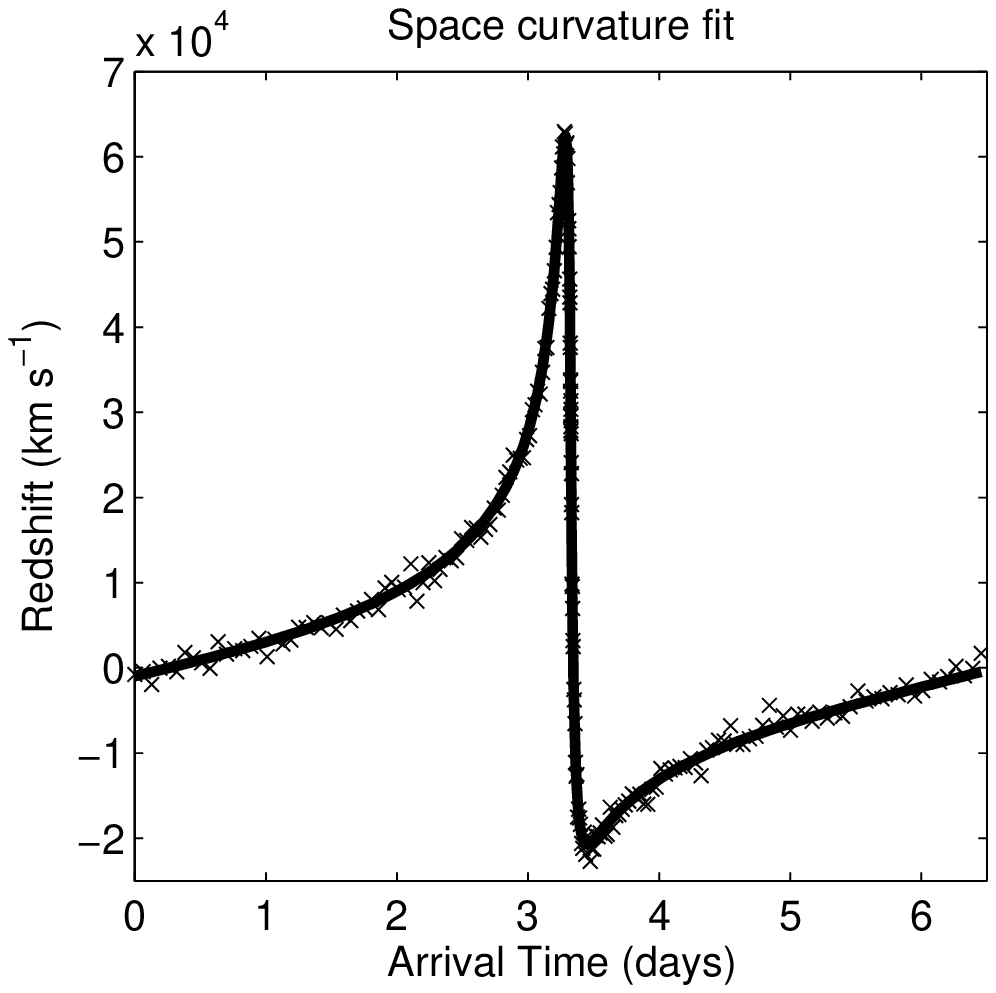}}
\caption{Our S2-like star has $a = a_{\rm S2}/100$. The mock redshift
  data in the above examples is generated all in the same way: 200
  data points are distributed along the complete orbit with a
  dispersion of $10^3 \kms$, gravitational time dilation and space
  curvature are all turned on. In the first panel, we fit with
  gravitational time dilation turned on, and space curvature turned
  off. For the second, we turn gravitational time dilation on, and for
  the third, we further turn space curvature on.  The mismatch between
  the fit and the simulated data is discernable in the first two
  panels.  Only in the last panel, when all the effects are included
  in the fit, is a satisfactory fit with $\chi^2_{\rm red} \approx 1$
  obtained.}\label{fig:results}
 \end{figure}

The redshift curve of a galactic centre star is dependent on a handful
of parameters. These include the Keplerian elements, the black hole
mass, as well as discrete parameters which toggle the post-Newtonian
terms. We proceed as follows. In order to put upper bounds on the
spectral resolution required to detect the post-Newtonian effects, we
generate mock spectral data consisting of 200 data points with a
chosen dispersion, using relativistic terms up to
$\mathcal{O}\left(v^3\right)$ for both the null and timelike metrics,
and determine whether or not we are able recover the parameter values
by fitting with these effects turned off. For illustrative purposes,
we consider an exaggerated S2 orbit. For the semi-major axis, we take
$a = a_{\rm S2}/100$. In doing so, $v=10v_{\rm S2}$ --- the classical
contribution to the redshift is raised 10-fold. The redshift due to
gravitational time dilation, which enters the dynamics at
$\mathcal{O}\left(v^2\right)$ is increased 100-fold. The space
curvature redshift contribution, entering at
$\mathcal{O}\left(v^3\right)$, is enhanced
1000-fold. Figure~\ref{fig:results} shows the results of the fitting
procedure. The classical fit manages a $\chi^2_{\rm red} = 4.88$, the
time dilation fit $\chi^2_{\rm red} = 2.68$, and the space curvature
fit $\chi^2_{\rm red} = 1.07$. Hence, a spectral dispersion of $10^3
\kms$ suffices for clear visual and numerical distinction of these
relativistic effects. In undoing our exaggeration of the orbit, we
argue that a spectral dispersion of $\sim 10\kms$ would allow for a
clear detection of gravitational redshift for the real S2, and $\sim
1\kms$ would yield a test for space curvature.

\section*{References}

\begin{enumerate}
\item
R.~Ang\'elil and P.~Saha.
\newblock {Relativistic redshift effects and the Galactic-center stars.}
\newblock {\em The Astrophysical Journal}, In Press, arXiv 1001.1957.

\item
S.~Zucker, T.~Alexander, S.~Gillessen, F.~Eisenhauer, and R.~Genzel.
\newblock Probing post-newtonian physics near the galactic black hole with
  stellar redshift measurements.
\newblock {\em The Astrophysical Journal Letters}, 639(1):L21--L24, 2006.

\item
R.~Kannan and P.~Saha.
\newblock {On post-Newtonian orbits and the Galactic-center stars}.
\newblock {\em Astrophysical Journal 690 (2009) 1553-1557}, September 2008.
\end{enumerate}
\end{document}